\documentclass[twocolumn,prl,tightenlines,superscriptaddress]{revtex4-1}

\usepackage{hyperref}
\usepackage{amsmath,amssymb}
\usepackage[utf8]{inputenc} 			
\usepackage[T1]{fontenc}				
\usepackage{graphicx}
\usepackage{float}
\usepackage[usenames,dvipsnames]{xcolor}
\usepackage{tikz}

\newcommand{\eps}{\varepsilon}

\renewcommand{\div}{\nabla\cdot}
\renewcommand{\vec}[1]{\mathbf{#1}}
\newcommand{\Gvec}[1]{\boldsymbol{#1}}

\newcommand*\colvec[3][]{
    \begin{pmatrix}\ifx\relax#1\relax\else#1\\\fi#2\\#3\end{pmatrix}
}

\usepackage{stmaryrd}

\begin{document}

\title{Metastability of Constant-Density Flocks}

\author{Marc Besse}
\affiliation{Sorbonne Universit\'e, CNRS, Laboratoire de Physique Th\'eorique de la Mati\`ere Condens\'ee, 75005 Paris, France}

\author{Hugues Chat\'{e}}
\affiliation{Service de Physique de l'Etat Condens\'e, CEA, CNRS Universit\'e Paris-Saclay, CEA-Saclay, 91191 Gif-sur-Yvette, France}
\affiliation{Computational Science Research Center, Beijing 100094, China}
\affiliation{Sorbonne Universit\'e, CNRS, Laboratoire de Physique Th\'eorique de la Mati\`ere Condens\'ee, 75005 Paris, France}

\author{Alexandre Solon}
\affiliation{Sorbonne Universit\'e, CNRS, Laboratoire de Physique Th\'eorique de la Mati\`ere Condens\'ee, 75005 Paris, France}

\date{\today}
 
\begin{abstract}
We study numerically the Toner-Tu field theory where the density field is maintained
constant, a limit case of ``Malthusian'' flocks for which the asymptotic scaling of correlation functions
in the ordered phase is known exactly. 
While we confirm these scaling laws, 
we also show that such constant-density flocks are metastable to the nucleation 
of a specific defect configuration, and are replaced by a globally disordered phase
consisting of asters surrounded by shock lines that
constantly evolves and remodels itself.  
We demonstrate that the main source of disorder lies along 
shock lines, rendering this active foam fundamentally different from the
corresponding equilibrium system.
We thus show that in the context of active matter also, a result obtained at all orders of
perturbation theory can be superseded by non-perturbative effects, calling
for a different approach.
\end{abstract}

\maketitle

Understanding simple models and theories of collective motion emerging from spontaneous symmetry breaking 
has spurred the field of active matter, and largely continues to animate it. 
Emblematic in this context are the Vicsek model~\cite{vicsek1995novel} and its field theoretical description, 
the Toner-Tu equations~\cite{toner1995long}.
In spite of a lot of progress, several aspects of their asymptotic behavior
are in fact still open problems. In particular,
the original derivation of exact scaling relations for the correlation functions of the
density and ordering (velocity) field in two dimensions (2D)~\cite{toner1998flocks}
was later shown to rely on unduly
neglecting some relevant terms~\cite{toner2012reanalysis}.
Unsurprisingly, numerical measurements of correlation
functions are inconsistent with the original
prediction~\cite{mahault2019TT}.

To make analytical progress, Toner and collaborators looked for simpler situations where
the couplings between the conserved density field and the ordering field, which are at the origin 
of the difficulties mentioned above, are qualitatively modified. One such option is to tame
the density field by considering ``Malthusian flocks'' in which particles die
and reproduce on a fast scale~\cite{toner2012birth,chen2020moving,chen2020universality}.
Another is to introduce an incompressibility constraint~\cite{chen2015critical,chen2018incompressible}.
In these two cases, remarkably, the scaling
laws governing fluctuations of the ordered phase can be calculated at all orders of
perturbation theory (in 2D for Malthusian flocks and any larger dimension for incompressible ones), 
providing rare examples of exact results about nontrivial out-of-equilibrium systems.

Apart from these analytical results, incompressible and Malthusian flocks
have received little attention. Rana and Perlekar have studied the coarsening to the ordered phase
of deterministic incompressible Toner-Tu equations~\cite{rana2020coarsening,rana2022phase}.
Two recent preprints on Malthusian flocks approached the
order-disorder transition. A one-loop dynamical renormalization
group approach~\cite{di_carlo_evidence_2022} suggests a
fluctuation-induced first order transition, like in the standard
compressible case \cite{martin_fluctuation-induced_2021}. 
A numerical study of a particle-based model concludes to a
crossover from first- to second-order behavior~\cite{mishra_active_2022}. 

Embarking on a numerical check of the exact results obtained by Toner {\it et al.} for the 
fluctuating ordered phase
could appear as a waste of time.  This is nevertheless what we did for Malthusian
flocks, motivated mostly by a string of recent results that
demonstrated the fragility of (usual, compressible) flocks to a series
of arbitrarily weak perturbations such as spatial
anisotropy~\cite{solon2022susceptibility}, quenched and chirality
disorder~\cite{toner2018swarming,toner2018hydrodynamic,yu2021breakdown,ventejou2021susceptibility},
and even one small fixed object~\cite{codina2022small}.  More
generally, results obtained at all orders in perturbation theory do
not necessarily mean they are the ultimate answer, since
non-perturbative effects can always arise. 
Famous examples include the KPZ
equation~\cite{wiese_perturbation_1998} and the Potts model in $6-\eps$
dimensions~\cite{amit_renormalization_1976,priest_critical_1976}.

Here we report on 2D numerical simulations of the Toner-Tu field
theory in which the density field is maintained strictly constant, a
limit case of Malthusian flocks. We find that while the predicted
scaling laws are indeed obeyed by the homogeneous flocking phase, this
phase is in fact always metastable to the nucleation of a specific
defect configuration, leaving eventually a globally disordered
cellular structure consisting of asters surrounded by shock lines that
constantly evolves and remodels itself.  The extended nature of these
shock lines, across which the orientation of the field changes
rapidly, renders this active foam fundamentally different from its
equilibrium equivalent, the disordered phase of the XY model, well
characterized in terms of a gas of pointwise topological defects. We
indeed show that the main source of disorder lies along and at the
intersection of shock lines.

Our starting point is the following minimal active field theory with advection and
alignment governing a single polarity/velocity field $\vec v$: 
\begin{equation}
  \label{eq:full-eq-adim}
  \partial_t \vec v+\lambda(\vec v\cdot \nabla)\vec v=\nabla^2\vec v+(a-|\vec v|^2)\vec v +\sqrt{2\eta}\,\Gvec\xi \,,
\end{equation}
where $\Gvec\xi$ is a delta-correlated, unit-variance, zero-mean Gaussian white noise 
and two coefficients have been set to unity without loss of
generality, leaving the three parameters $a$, $\lambda$ and the noise
intensity $\eta$. One more parameter could be set to unity but we
retain these three to explore the different limits of the
model. Equation~(\ref{eq:full-eq-adim}) can be obtained from the field
theory for Malthusian flocks by integrating out the fast density
field, as done in ~\cite{toner2012birth}.  It can also be considered
as the standard Toner-Tu theory where the density is kept constant,
without imposing incompressibility. Note that
Eq.~(\ref{eq:full-eq-adim}) is invariant under the transformation
$\vec v\to -\vec v$ and $\lambda\to -\lambda$.  For simplicity, we do
not include the two other possible advection terms
$(\div \vec v)\vec v$ and $\nabla(|\vec v|^2)$ for which we checked
that the same type of behaviour is observed.  In the following, we
integrate Eq.~(\ref{eq:full-eq-adim}) in 2D square domains of linear size $L$ 
using a pseudo-spectral
method with Euler explicit time stepping and antialiasing. In the rest of the paper,
the spatial resolution is set to $dx=2$ and the time resolution to
$dt=0.1$, unless otherwise explicitly stated.

In 2D, Eq.~(\ref{eq:full-eq-adim}) for $\lambda=0$ reduces to the
relaxational dynamics of a field theory describing the XY model. 
We thus expect a Berezinskii–Kosterlitz–Thouless
transition from disorder to quasi-long-range order (QLRO) when, say, increasing
$a$~\cite{berezinskii_destruction_1971,kosterlitz_ordering_1973}.
On the other hand the $\lambda$ advection term controls the activity
level and Toner {\it et al.} predict true long-range order (LRO) for
large-enough $a$ whenever $\lambda\ne 0$.

Simulating Eq.~(\ref{eq:full-eq-adim}) with a
large enough mass $a$ and starting from an ordered initial condition,
the system indeed settles in a symmetry-broken state with a non-zero mean velocity. 
To distinguish between true long-range order and the
quasi-long-range order expected in the passive case, we compute the order parameter
$\bar v=\langle |\langle \vec v\rangle_{\vec x}| \rangle_t$ for various system sizes $L$
($\langle \cdot \rangle_{\vec x}$ and
$\langle \cdot \rangle_t$ denote respectively space and time averages). 

\begin{figure}[tbp!]
  \centering
    \includegraphics[width=0.48\linewidth]{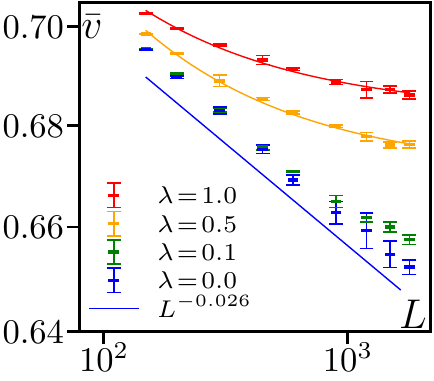}
    \includegraphics[width=0.48\linewidth]{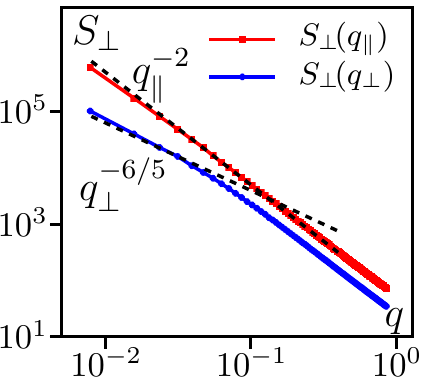}
    \caption{{\bf Left:} $\bar v(L)$ for different $\lambda$ values
      ($a=0.7, \eta=0.5$ and initial ordered state).
      Error bars correspond to two standard errors of
      the mean computed over 5 independent samples. 
      For large $\lambda$ values, $\bar{v}(L)$ is
      reasonably well fitted by
      $\bar{v}-\bar{v}_{\infty} = A L^{-\omega}$ with
      $\omega=2/3$  
      ($\bar{v}_{\infty}=0.67, A=0.77$, for $\lambda=0.5$, orange curve, and 
      $\bar{v}_{\infty}=0.68, A=0.57$, for $\lambda=1.0$, red curve). 
      {\bf Right:} Variation of structure factor of $v_\bot$ with
      $q_\parallel$ for $q_\bot=0$ (red squares) and with $q_\bot$ for
      $q_\parallel=0$
      ($a=5$, $\lambda=1$, $\eta=0.5$, $L=800$).
      Dashed lines show the scaling predicted at
      small wave vectors~\cite{toner2012birth}, which arises only beyond a crossover scale for $S_\bot(q_\bot)$.}
\label{fig:ordered}
\end{figure}

In the passive limit $\lambda=0$, we see $\bar v(L)\sim L^{-\theta}$, the signature of QLRO 
(a typical case is shown in Fig.~\ref{fig:ordered}(a)).
The exponent $\theta$ varies continuously with $a$ (not shown). Using its expected value $\theta=\tfrac{1}{8}$
at the BKT transition, 
we estimate the latter to happen at $a\simeq 0.38$.

In the active case with not too small $\lambda$, in contrast, $\bar v(L)$ 
decays slower than a power law, suggesting a finite value $\bar v_\infty$ in the infinite size limit.
As expected, 
our data is reasonably well fitted by $\bar v -\bar v_\infty \sim L^{-\omega}$, indicating LRO.
Decreasing $\lambda$, this behavior is only observed beyond a
crossover scale that seems to diverge when $\lambda\to 0$
(Fig.~\ref{fig:ordered}(a)).

To investigate further the active ordered phase, we computed the scaling of
velocity fluctuations via the structure factor, {\it i.e.} the Fourier transformed
equal-time spatial correlation function
\begin{equation}
  \label{eq:Sq}
 S_\bot(\vec q)=\langle |v_\bot(\vec q,t)|^2\rangle_t 
\end{equation}
where $v_\bot(\vec q)$ is the velocity orthogonal to the direction of global order at
wavevector $\vec q=(q_\bot, q_\|)$
\footnote{One has to be careful in
taking the time average in Eq.~(\ref{eq:Sq}) because the direction of
global order drifts slowly.
In practice, following Ref.~\cite{mahault2019TT}, we apply an external field
$\vec h$ large enough to pin the global direction of order
but small enough that it does not affect the correlation functions on
the range of $q$-values displayed.}. 
Our results, shown in Fig.~\ref{fig:ordered}(b), indicate anisotropic scaling in good quantitative agreement with
the predictions of Toner {\it et al.} in the small wavenumber limit:
$S_\bot(q_\parallel)\sim |q_\parallel|^{-2}$ if
$q_\parallel\gg q_\bot$ and $S_\bot(q_\bot)\sim |q_\bot|^{-6/5}$ if
$q_\bot\gg q_\parallel$~\cite{toner2012birth}.

The agreement between our measurements and the theoretical prediction
is hardly surprising. 
However, as we show now, this is far from being the whole story, as more complex
structures, to which the renormalization group approach of
Ref.~\cite{toner2012birth} is blind, appear.

\begin{figure*}
  \centering
 \includegraphics[width=\linewidth]{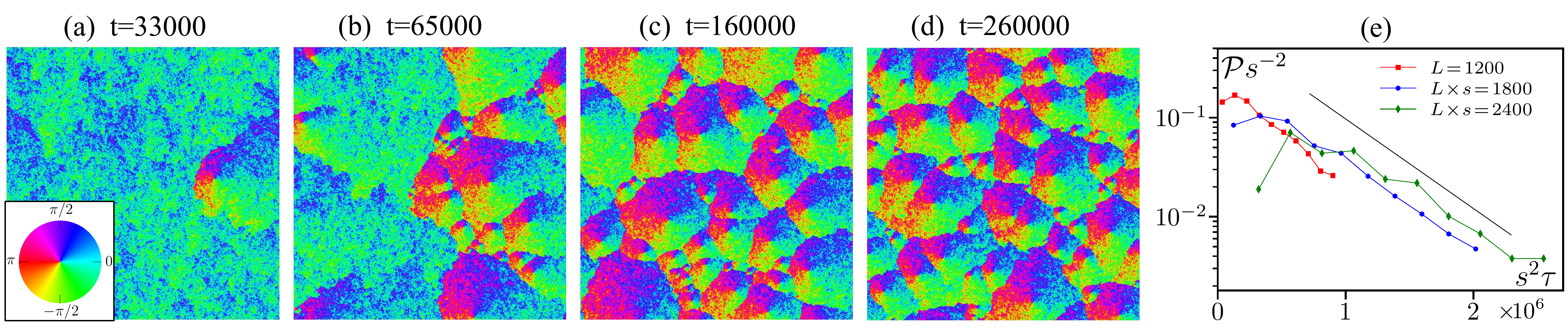}
  \caption{{\bf (a-d):} Snapshots taken during a run starting from an ordered state showing the nucleation
  of a first defect and the following
    evolution ($a=0.48, \lambda=1.0, \eta=0.5$, $L=3600$). 
    {\bf (e):} Probability
    distribution ${\cal P}$ of the lifetime of the ordered phase $\tau$, defined
    at the first time for which the nucleation of an aster decreases
    $\bar{v}$ by more than
    20\% ($a=0.45,\lambda=1.0,\eta=0.5$). 
    Data obtained at 3 different $L$ values, rescaled by a factor $s^2$ proportional to $L^2$.}
\label{fig:nucleation}
\end{figure*}

For not too large $a$ values, in a large-enough system, the ordered
phase described above is easily observed to break down spontaneously:
fluctuations can lead to the emergence of a specific local
configuration made of an aster surrounded by a semi-circular shock
line (Fig.~\ref{fig:nucleation}(a)). The aster is a point around which
polarity is arranged radially, and where a $+1$ topological defect is
thus located.  Across the shock line, polarity varies rapidly. It
actually embeds a $-1$ defect as we will see later.  In large systems,
this initial nucleation is followed by others located elsewhere,
and/or the emergence of new asters (and their accompanying shock
lines) along the shock line of the initial aster
(Fig.~\ref{fig:nucleation}(b)).  Eventually, this proliferation
process stops and a globally-disordered, dynamical state is reached,
with a well-defined average number of asters and a web of shock lines
surrounding each of them (Fig.~\ref{fig:nucleation}(c,d) and Movie~1
in \cite{SUPP}).  In this steady state, new asters are constantly
generated near shock lines and dominantly near their intersections,
while existing asters have their position and size fluctuate in time
until one of their surrounding shock line meets their center, at which
point they disappear.

We studied the statistics of the nucleation process of the first aster from the ordered phase, 
following ordered initial conditions. 
Figure~\ref{fig:nucleation}(e) shows that the lifetime $\tau$ of the ordered phase, 
i.e. the time it takes for nucleation events to destroy order, is distributed exponentially, 
with a characteristic time that scales roughly like $1/L^2$, the inverse system size.
In other words, the nucleation rate is proportional to $L^2$.
This confirms that the emergence of asters and their surrounding shock line is essentially 
a local nucleation process, even though the ordered phase possesses built-in long-range correlations.

\begin{figure}[b!]
  \centering
    \includegraphics[width=0.43\linewidth, height=0.69\linewidth]{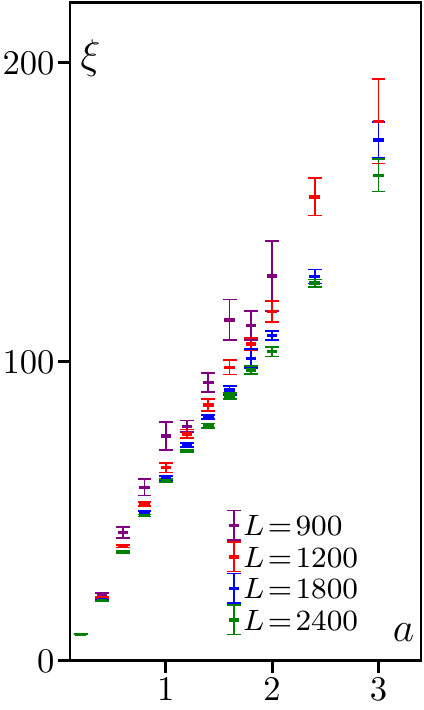}
    \includegraphics[width=0.55\linewidth]{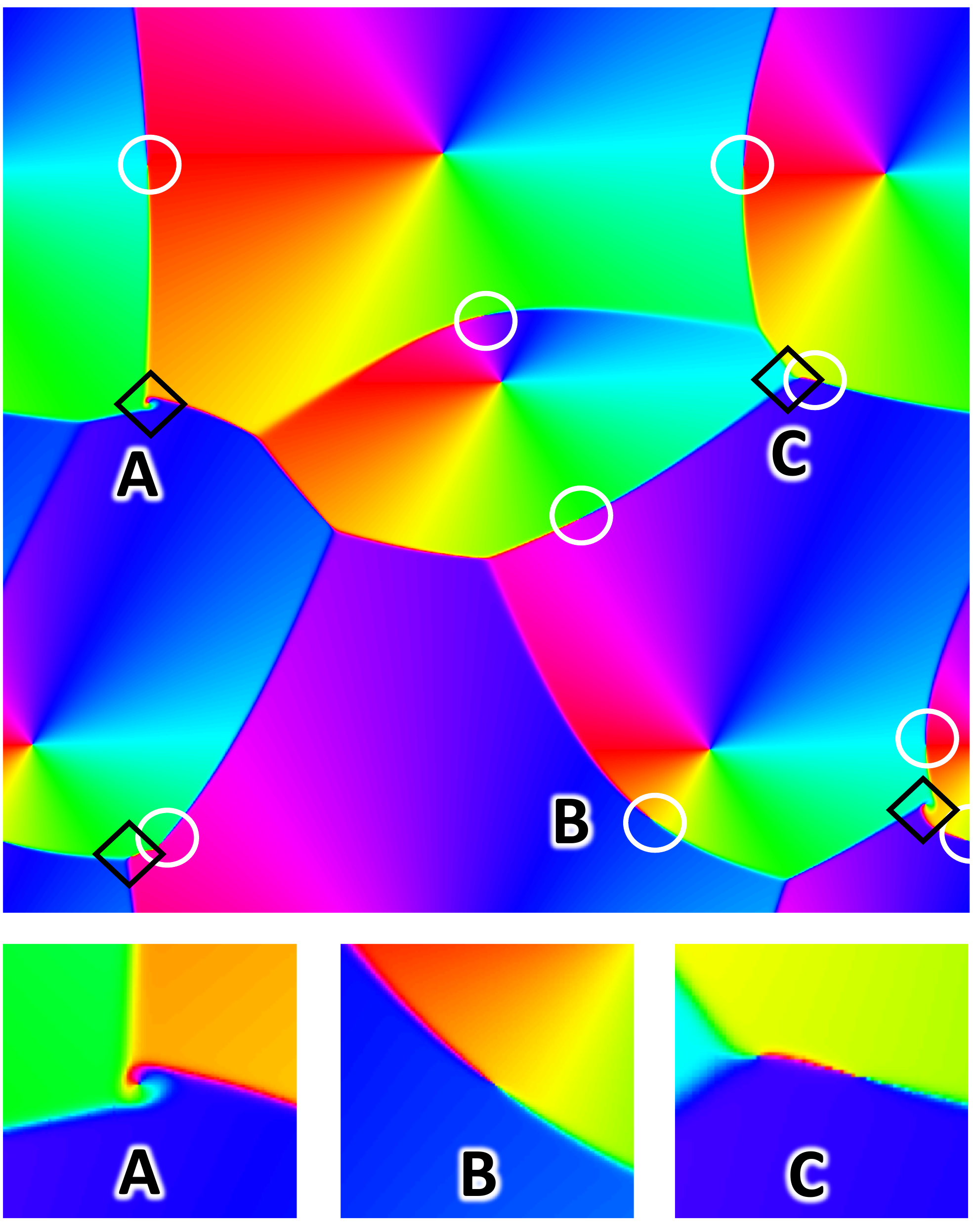}
    \caption{{\bf Left:} Variation of correlation length $\xi$ with
      $a$ for different sizes $L$. Data taken in the active foam
      steady state ($\lambda=1.0, \eta=0.5$).  Error bars correspond
      to two standard errors of the mean, computed over $10$
      independent time intervals.  At a given $a$ value, small size
      data may not be reliable since $\xi$ can then be of the order of
      $L$.  At $L=900$ in particular, data for $a>2.0$ are not shown
      since only very few or even no asters are present.  {\bf Right:}
      Typical configuration taken during coarsening in the
      deterministic limit $\eta=0$ ($a=1$, $\lambda=1.0$, $L=900$,
      $dx=0.5$, $dt=0.01$, colors as in Fig.~\ref{fig:nucleation}(a),
      from Movie~3 in \cite{SUPP}).  Apart from the 5 clearly visible
      asters, which have $+1$ topological charge, 9 shock-line
      embedded $-1$ defects are present (white circles), as well as 4
      vortex-like $+1$ defects present at some shock line vertices
      (black diamonds).  Labels A, B, C in the main panel point to the
      defects shown more clearly in the small lower panels.}
\label{fig:defects}
\end{figure}

Replicating this study at various parameter values, and in particular at large values of $a$, 
quickly becomes numerically prohibitive because $\tau$ can then take very large values.
The data at hand does not suggest the existence of a threshold beyond which nucleation would become impossible.
This is corroborated by a direct study of the dynamic active foam. 
This steady state is easily reached from disordered initial conditions after a typically
short coarsening transient (cf. Movie~2 in \cite{SUPP}). 
This allows to study it at parameter values for which the spontaneous nucleation of an 
aster from the ordered phase would take unreasonably long times.
The active foam can be characterized by its average number of asters, but we preferred to measure
the more robust correlation length $\xi$ extracted from the radially averaged structure 
factor~\footnote{$\xi=2\pi\!\int\! dq\, S(q)/ \!\int \!  dq\, q S(q)$ with $q=|\vec q|$.}. 
At fixed parameters, $\xi$ converges to a well-defined asymptotic value when the system size $L$ is increased: 
the active foam is self-averaging, and reliable estimates of its intrinsic correlation length are obtained as soon as 
$L\gg \xi$.
At fixed activity $\lambda$, we find that $\xi$ increases with increasing $a$, 
but this growth is modest, possibly linear with $a$ (Fig.~\ref{fig:defects}, left). Contrary to the passive case where the BKT transition
at $a\simeq 0.38$ marks the fast divergence of correlations, we do not see any sign of a transition beyond which 
the ordered phase would remain stable. 
Extrapolating our numerical results, we conclude that the ordered phase is metastable 
for any non-zero $\lambda$ and positive $a$~\footnote{Ignoring a possible renormalized value for the $a=0$
mean-field threshold}. 

To better understand why the active regime is so different from the
passive case, we now turn to a more detailed study of the objects at
play in the dominating active foam phase, i.e. asters and shock lines.

Asters have been reported in models of the self-organization of microtubules 
and molecular motors~\cite{lee2001macroscopic,sankararaman2004self-organized,aranson2005pattern,aranson2006theory},
but also in
variants of the Toner-Tu equations~\cite{gopinath2012dynamical,gowrishankar2016nonequilibrium,husain2017emergent,elgeti2011defect,sankararaman2004self-organized}, as well
as in active gel theory~\cite{kruse2004asters,kruse2005generic}.
All these systems are more complex than ours, including a density and/or a motor concentration field.
To our best knowledge, only Vafa \cite{vafa2020defect} studied defects in the simple 
Equation~\eqref{eq:full-eq-adim} of interest here, concluding that asters are the most stable $+1$ 
defects~\footnote{Note that asters are pointing outward with $\lambda>0$, and inward when with $\lambda<0$}. 
This result is in agreement with our observations of the active foam, where asters are quite passive regions,
and most of the remaining activity occurs at the shock lines and in particular at their intersection 
(cf. Movies~1 and 2 in \cite{SUPP}).

Active foam configurations are best understood in the deterministic version of Eq.~(\ref{eq:full-eq-adim}), 
either by switching off the noise from a given configuration, 
or by watching the slow coarsening following disordered initial conditions 
(cf. Movie~3 in \cite{SUPP})~\footnote{In the deterministic case,
we believe, in agreement with Vafa \cite{vafa2020defect}, that coarsening should proceed
all the way to the ordered phase; in practice finite numerical resolution may pin the system in some near final configuration with few asters.}. 
The right panel of Fig~\ref{fig:defects} shows a late configuration comprising a few remaining 
asters and their surrounding shock lines, extracted from this coarsening process. 
Close inspection shows that shock lines are extended objects across which the phase varies rapidly but not discontinuously.
Given the existence of $+1$ defects (at the centers of asters), $-1$ defects must be present. 
In the 5-aster configuration under scrutiny, 9 defects with topological charge $-1$ are found 
embedded in the shock lines (white circles, zoom B), 
typically located at the locations closest to the aster centers, where the phase jump is $\pm\pi$. 
The $-2\pi$ circulation around these defects is mostly due to two phase jumps occurring when crossing the shock line.  
This suggests that the important structures are the extended shock lines and not so much the 
pointwise location of the $-1$ defect.
Given that the global topological charge must be zero, our inspection is not complete. 
Indeed one can locate 4 other $+1$ vortex-like defects
typically located at shock lines vertices (black diamonds and zooms A, C). 
Note finally that the shock line vertices are rather unstable dynamically, even in the absence of noise 
(cf. Movie~3 \cite{SUPP}).

\begin{figure}[b!]
  \centering
  \includegraphics[width=1\linewidth]{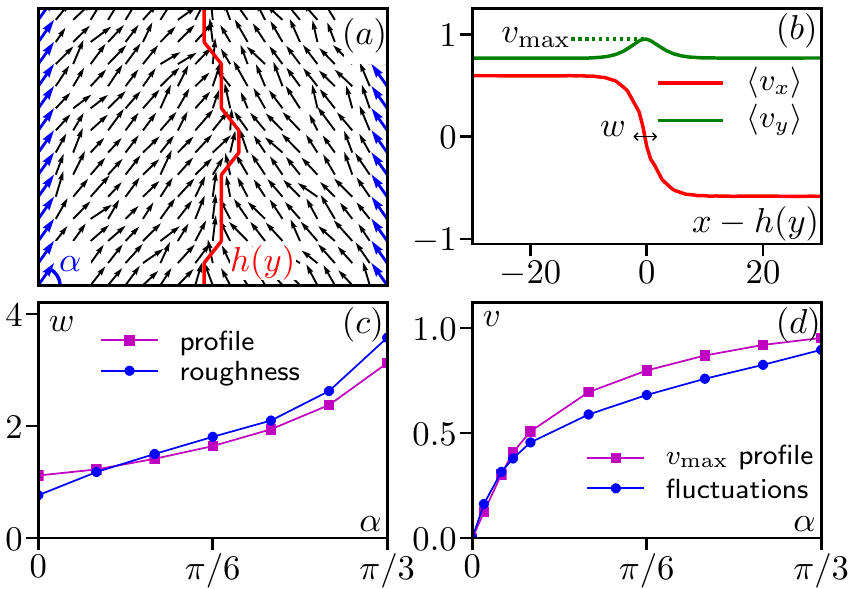}
  \caption{Study of the interface separating two half domains with different bulk orientation.
  (a): Typical snapshot showing the interface $h(y)$ (red line) defined by the points along $x$ where
    $v_x$  changes sign. The velocity is
    fixed on the left and right boundaries (blue arrows),
    at an angle $\alpha=\pi/3$ with respect to the horizontal  
    ($a=1, \lambda=1, \eta=0.1$, $300\times 100$ system, periodic boundary condition in $y$). 
    (b): Mean horizontal profiles of $v_x$ and $v_y$ 
    measured relative to the interface position $h(y)$, averaged over $y$ and time
    (same parameters as in (a).
 (c): variation with $\alpha$ of the width of the
    interfacial region, measured either by fitting the average profile
    $\langle v_x \rangle$ (as in (b)) by a $\tanh$ function (magenta
    squares) or as the roughness of the interface
    $w=\sqrt{\langle \langle h^2\rangle_y-\langle
      h\rangle_y^2\rangle_t}$ (blue circles). System size $100\times 100$, other parameters as in (a,b).
  (d): same as (c) but showing the propagation speed
    of fluctuations along the interface obtained either as the maximum
    of $\langle v_y\rangle$ in (b) (magenta squares) or from the
    space-time Fourier transformed correlation function
    $\langle |h(q_y,\omega)^2|\rangle_t$ as the position of the peaks
    $\omega=v q_y$ at small wave-vector $q_y$. 
    }
\label{fig:interfaces}
\end{figure}

Coming back to the noisy case, the radially organized aster regions
appear rather stable ---in agreement with Vafa's
findings~\cite{vafa2020defect}---, and the strongest fluctuations are
found along the shock lines (Movies~1 and 2 in \cite{SUPP}).  We have
studied the behavior of these fluctuations in specially prepared
systems with a single shock line separating two large ordered
homogeneous domains whose orientations differ by some angle
$\pi-2 \alpha$ (Fig.~\ref{fig:interfaces}(a)). The profile of the
interface is shown in Fig.~\ref{fig:interfaces}(b) and is found to be
close to its deterministic shape. Under the effect of noise the
position of the interface fluctuates leading to a roughness which we
find to increase monotonically with $\alpha$ in the same way as the
width of the mean profile (Fig.~\ref{fig:interfaces}(c)). For
$\alpha>0$, parity symmetry is broken along the interface, and
fluctuations on the shock line are advected along the direction given
by the sign of $\langle v_y\rangle$.  We find that the advection speed
of fluctuations increases with $\alpha$ in the same way as the maximum
of $\langle v_y\rangle$ (Fig.~\ref{fig:interfaces}(d)).

In typical active foam configurations such as those shown in
Figs.~\ref{fig:nucleation}(a-d),\ref{fig:defects}(b), the local
orientation difference across a shock line varies with the position
along the line. In the notation introduced above, near the
intersection with the line joining the two aster centers (where a $-1$
defect typically sits), the phase jump corresponds to $\alpha\simeq 0$
while going away from this point means increasing $\alpha$.  Using the
results obtained in Fig.~\ref{fig:interfaces} with constant-$\alpha$
shock lines, we can now understand why fluctuations are growing and
are advected faster and faster when going away from the
$\alpha\simeq 0$ point, and why they are maximal at shock line
intersections, making these regions the most unstable ones where new
asters are created.  Thus the mechanisms at the origin of local
dynamics maintaining the active foam in a steady state are intimately
linked to the extended nature of shock lines and are very different
from the unbinding of nucleated defect pairs characteristic of the
disordered phase of the XY model.

In conclusion, we have shown numerically that fluctuations in the
ordered phase of the 2D constant-density Toner-Tu theory obey the
scaling laws predicted by a perturbative renormalization group
analysis, but we have also found that these flocks are metastable to
an ever-evolving active foam state made of asters surrounded by a
network of shock lines.  This constitutes an instance, in the context
of active matter, where a result obtained at all orders of
perturbation theory is superseded by non-perturbative effects, calling
for a different approach.

We have demonstrated that the shock lines separating asters are
crucial to amplify and advect perturbations, so that the vertices of
the network they form are the most susceptible regions of space where
new asters are created.  The topology and dynamics of the structures
at play are thus qualitatively different from the binding/unbinding of
pairs of point defects that rule the fate of order in the passive case
of the XY model. Our findings are reminiscent of the recent study of coarsening 
in the compressible case, where a network of ``domain walls'' was put forward
but no asters are present~\cite{chardac2021topology}. Coarsening was
also studied in the incompressible case, but there the dominating structures
are standard vortices and anti-vortices \cite{rana2020coarsening}. 
The objects at play in defected flocking systems can thus vary qualitatively, an observation that
calls for further study. 
In fact our active foam may be closest to the 
residual chaos present in some regimes of the 2D complex
Ginzburg-Landau equation where stable spiral wave domains are
surrounded by a network of shock lines at which the traveling waves
emitted by spirals meet~\cite{brito2003vortex}. 

The field theory studied here is a limit case of that
for Malthusian flocks, where a fast but finite timescale regulates the
density field. It would thus be interesting to see how our results
translate to this more general setting, including in microscopic,
particle-level models.  Future work should also investigate the
possibility of similar ``non-perturbative metastability'' of long-range
orientational order in other important models and field theories of
active matter both in 2D and 3D.

Finally, our findings could be of relevance in some real systems in
spite of the simplicity of the framework we considered.  Prime
candidates are found in cytoskeletal active matter, {\it i.e.} {\it in
  vitro} mixtures of (mostly) biofilaments and molecular motors, for
which the formation of asters have been
reported~\cite{nedelec1997self-organization,surrey2001physical,backouche2006active,hentrich2010microtubule,phuong2014spatial,stam2017filament,khetan2021self-organized,berezney2022extensile,lemma2022active}

\acknowledgements We thank Matthieu Tissier for insightful discussions
and Francesco Ginelli, Benoît Mahault and Xia-qing Shi for their
comments on the manuscript.  HC's work was supported in part by ANR
project NeqFluids grant ANR-18-CE92-0019 and the National Natural
Science Foundation of China (Grant No. 11635002).

\bibliography{Biblio-current.bib}

\end{document}